\documentclass{article}

\usepackage[utf8]{inputenc} % allow utf-8 input
\usepackage[T1]{fontenc}    % use 8-bit T1 fonts
\usepackage[hyphens]{url}
\usepackage[breaklinks=true]{hyperref}
\usepackage{booktabs}       % professional-quality tables
\usepackage{amsfonts}       % blackboard math symbols
\usepackage{amsmath} 
\usepackage{nicefrac}       % compact symbols for 1/2, etc.
\usepackage{microtype}      % microtypography
\usepackage{lipsum}
\usepackage{graphicx}       % graphics
\graphicspath{{media/}}     % organize your images and other figures under media/ folder

\hypersetup{%
    hypertexnames=false,
    colorlinks=true,
    allcolors=blue,
    unicode=true
}
\title{Purported quantitative support for multiple introductions of SARS-CoV-2 into humans is an artefact of an imbalanced hypothesis testing framework}

\author{
  \textbf{Angus McCowan}\thanks{\texttt{awmccowan@fastmail.com}}\\
}
\date{}
\begin{document}
\maketitle

\begin{abstract}
A prominent report claimed substantial support for two introductions of SARS-CoV-2 into humans using a calculation that combined phylodynamic inferences and epidemic models. Inspection of the calculation identifies an imbalance in the hypothesis testing framework that confounds this result; the single-introduction model was tested against more stringent conditions than the two-introduction model. Here, I show that when the two-introduction model is tested against the same conditions, the support disappears.
\end{abstract}

Understanding the number of SARS-CoV-2 introductions into humans is useful for estimating the genetic diversity in the source, and is also relevant to understanding possible links between the source and particular infection clusters.

A prominent report \cite{doi:10.1126/science.abp8337} reviewed genomic data from early in the epidemic and concluded that sequences outside of two main lineages were unreliable. The authors of the report then hypothesized that the two lineages arose from separate introductions into humans, and quantified support for this hypothesis with a Bayes factor ($\mathrm{BF}$) that compared the plausibility of their curated genomic data ($\mathbf{Y}$) arising from two successful introductions ($I_2$) versus one ($I_1$),

\begin{equation}
    \mathrm{BF} = \frac{P(\mathbf{Y} | I_2)}{P(\mathbf{Y} | I_1)}.
    \label{eq:bayes_factor}
\end{equation}
 
The published Bayes factor was 60, with which the authors claimed strong support for multiple introductions. Subsequent reproduction efforts identified three errors in the calculation: a syntax error caused undercounting of one-introduction likelihoods; an erroneous normalization disproportionately inflated the two-introduction likelihoods; and non-exclusive marginalization caused double-counting of two-introduction likelihoods. With a subsequent erratum \cite{doi:10.1126/science.adl0585}, the syntax error was corrected, the erroneous normalization removed and the marginalization made exclusive. As a result, the Bayes factor dropped to 4.

A Bayes factor of 4 should normally indicate substantial or moderate support. However, there is an imbalance in the conditioning of the likelihoods used to calculate Equation \ref{eq:bayes_factor}.

The imbalance is the focus of this investigation. I identify the imbalanced conditions, including a relative-clade size condition, an evolutionary separation condition, and a root shape condition. I then evaluate the effect of the relative-clade size condition, and explore the additional effect of the evolutionary separation and root shape conditions under minimal assumptions.

\section{Identifying the imbalance}
The authors evaluated the Bayes factor \ref{eq:bayes_factor} by synthesizing the results of two different analyses: Bayesian phylodynamic inference and forward epidemic simulations. 

The Bayesian phylodynamic inference used the genomic data $\mathbf{Y}$ of sequenced virus samples, including the dates of the samples, to reconstruct posterior distributions of the most recent common ancestor (${\text{MRCA}}$). Four candidates for the haplotype ($S_{\text{MRCA}}$) of the ${\text{MRCA}}$ were identified: $S_{A}$ (one of the two main lineages, where nucleotides at two specific positions were the same as in bat viruses), $S_{B}$ (the other main lineage, where nucleotides at both those positions were different), $S_{C}$ (an intermediate lineage, with one of the nucleotides that characterize $S_{A}$ and one of the nucleotides that characterize $S_{B}$), and $S_{T}$ (an intermediate lineage complementary  to $S_{C}$, with the other of the nucleotides that characterize $S_{A}$ and the other of the nucleotides that characterize $S_{B}$). Under the most plausible inference scheme, which used a synthetic recombinant ancestor (recCA) as  root, the authors reported posterior probabilities of $P(S_{\text{A}} | \mathbf{Y}) = 77\%$, $P(S_{\text{B}} | \mathbf{Y}) = 8\%$, $P(S_{\text{C}} | \mathbf{Y}) = 10\%$ and $P(S_{\text{T}} | \mathbf{Y}) = 4\%$. 

The Bayesian phylodynamic inference also suggested associations between ancestral haplotypes and mutation tree shapes. An $MRCA$ of either of the two main lineage haplotypes $S_{A}$ or $S_{B}$ was associated with one of the main lineages being rooted at the $MRCA$ and the other being at the end of a two mutation branch from the $MRCA$. An $MRCA$ of either of the two intermediate lineage haplotypes $S_{C}$ and $S_{T}$ was associated with each of the two main lineages being rooted at the end of respective one mutation branches from the $MRCA$.

The forward epidemic simulations were used to estimate likelihoods of these mutation tree shapes. Specifically, the authors considered two one-introduction topologies, $\tau_{1C}$ and $\tau_{2C}$, and one two-introduction topology, $(\tau_P, \tau_P)$. The authors' estimated likelihoods were $P((\tau_P,\tau_P) | I_2) = 22.6\%$, $P(\tau_{1C} | I_1) = 3.1\%$ and $P(\tau_{2C} | I_1) =  0.0\%$.

Thus, the Bayesian phylodynamic inference estimated the posterior probabilities of ancestral haplotypes given the genomic data ($P(S_{\text{MRCA}} | \mathbf{Y})$), while the epidemic simulations estimated likelihoods of mutation tree topologies from one or two introductions ($P(\tau | I)$).

In order to combine these results, the authors marginalized the Bayes factor  \ref{eq:bayes_factor} over the ancestral haplotypes $S_{\text{MRCA}}$ and the mutation tree topologies $\tau$, while assuming equal prior probabilities for ancestral haplotypes to derive

\begin{equation}
\mathrm{BF} \;=\; 
\frac{
  \displaystyle \sum_{S_{\text{MRCA}}} 
    P\bigl(S_{\text{MRCA}} | \mathbf{Y}\bigr) 
    \;\Biggl[\sum_{\tau} P\bigl(S_{\text{MRCA}} | \tau\bigr)\,P\bigl(\tau | I_2\bigr)\Biggr]
}{
  \displaystyle \sum_{S_{\text{MRCA}}} 
    P\bigl(S_{\text{MRCA}} | \mathbf{Y}\bigr)
    \;\Biggl[\sum_{\tau} P\bigl(S_{\text{MRCA}} | \tau\bigr)\,P\bigl(\tau | I_1\bigr)\Biggr]
}
\label{eq:bayes_factor_detailed}
\end{equation}

where

\begin{equation}
    P(S_{\text{MRCA}} | \tau) \propto 
    \begin{cases} 
        1 & \text{if } \tau \text{ and } S_{\text{MRCA}} \text{ are compatible} \\ 
        0 & \text{otherwise}
    \end{cases}
    \label{eq:compatibility}
\end{equation}

and

\begin{equation}
    \sum_{S_{\text{MRCA}}} P\bigl(S_{\text{MRCA}} | \tau\bigr) \;=\; 1 
\quad\text{for all \(\tau\).}
    \label{eq:unity}
\end{equation}

That is, the Bayes factor mixed the posterior probabilities of the ancestral haplotypes $P(S_{\text{MRCA}} | \mathbf{Y})$ with the likelihoods of the mutation tree topologies $P(\tau | I)$ through compatibility equations ($P(S_{\mathrm{MRCA}} | \tau)$), such that the likelihood of each topology was weighted  according to the average posterior probability of the ancestral haplotypes compatible with that topology. The compatibility may be summarized as follows:

\begin{itemize}
    \item $\tau_{2C}$ involves the two main lineages branching from an unobserved lineage, and was deemed compatible with $\{S_C,S_T\}$
    \item $\tau_{1C}$ involves one of the two main lineages branching from the other, and was deemed compatible with $\{S_A,S_B\}$
    \item $(\tau_P, \tau_P)$ did not distinguish the branching, and was deemed compatible with $\{S_A, S_B, S_C, S_T\}$  
\end{itemize}

With these compatibility statements, Equations \ref{eq:compatibility} and \ref{eq:unity}  evaluate to the compatibility equations

\begin{align}
P\bigl(S_{\mathrm{MRCA}} \mid \tau_{2C}\bigr)
  &=
  \begin{cases}
    \tfrac12, & \text{if } S_{\mathrm{MRCA}} \in \{S_C,\,S_T\},\\[6pt]
    0,        & \text{otherwise,}
  \end{cases}
  \label{eq:compatibility_2c} \\[10pt]
P\bigl(S_{\mathrm{MRCA}} \mid \tau_{1C}\bigr)
  &=
  \begin{cases}
    \tfrac12, & \text{if } S_{\mathrm{MRCA}} \in \{S_A,\,S_B\},\\[6pt]
    0,        & \text{otherwise,}
  \end{cases}
  \label{eq:compatibility_1c} \\[10pt]
P\bigl(S_{\mathrm{MRCA}} \mid (\tau_P, \tau_P)\bigr)
  &=
  \begin{cases}
    \tfrac14, & \text{if } S_{\mathrm{MRCA}} \in \{S_A,\,S_B,\,S_C,\,S_T\},\\[6pt]
    0,        & \text{otherwise.}
  \end{cases}
  \label{eq:compatibility_pp}
\end{align}

The imbalance lies in the conditioning of the likelihoods, $P(\tau | I)$, through the different requirements imposed on the different mutation tree topologies, $\tau$. All topologies were required to have all taxa within only two clades, with a polytomy of at least 100 lineages at the root of each, but the one-introduction topologies were tested against three additional conditions: 
\begin{itemize}
    \item a relative size condition - neither clade could comprise less than 30\% of all taxa;
    \item an evolutionary separation condition - the roots of each clade had to be separated from each other by at least two mutations; and
    \item a root shape condition - the one-introduction topologies were specified as either $\tau_{1C}$ (if one of the two clades was ancestral and the other derived), or $\tau_{2C}$ (if both clades were derived).
\end{itemize}

One-introduction virus trees that failed to satisfy the relative size and evolutionary separation conditions were excluded from the Bayes factor calculation. Equivalent two-introduction virus trees were not. 

One-introduction virus trees were divided by the root shape condition into specific topologies, each with the specific compatibility equations \ref{eq:compatibility_2c} and \ref{eq:compatibility_1c}. Two-introduction virus trees were not, but the marginalization gave them the compatibility equation \ref{eq:compatibility_pp}, which is the average of the one-introduction  compatibility equations \ref{eq:compatibility_2c} and \ref{eq:compatibility_1c}. Thus, the weighting of the one-introduction likelihoods was distributed according to the root shape, but that of the two-introduction likelihood was simply averaged. 

The Bayes factor cannot provide meaningful support for the two-introduction hypothesis unless the effect of this imbalance was negligible.

Further minor imbalance arises in the filtering of early samples and short-lived basal lineages. For the one-introduction model, these are determined, respectively, by the first hospitalization and MRCA of the entire epidemic. For the two-introduction model, these are determined, respectively, by the first hospitalization and MRCA of each clade. Since this filtering involved the human alertness to the epidemic (the start of sampling and detection), the filtering of the two-introduction model should be based on the first hospitalization and MRCA of the entire epidemic, rather than filtering each clade separately based on their respective first hospitalization and MRCA.

\section{Evaluating the effect of the relative size condition}

I first reproduced the authors' epidemic simulations with a hundredfold increase in sample size, to generate 110,000 simulated successful epidemics. Then, following the authors' protocols, I simulated sequencing, coalescence and evolution to generate a mutation tree from each successful epidemic, and counted the proportion with topologies $\tau_{1C}$ and $\tau_{2C}$. I did not use the stable coalescence because it would prune short-lived basal lineages, to the benefit of the one-introduction model but not the two-introduction model. The resulting one-introduction likelihoods were $P(\tau_{1C} | I_1) = 3.1\%$ and $P(\tau_{2C} | I_1) = 0.1\%$, reproducing the results of the authors within sampling error.

In order to evaluate the effect of the relative size condition, I then combined successful epidemics and simulated the same sequencing, coalescence and evolution to generate a mutation tree from each combined epidemic. In order to maximize the likelihood of the two introductions satisfying the relative size condition, I made the introductions simultaneous. I then counted the proportion with the topology $(\tau_P,\,\tau_P)$, where the separately introduced lineages each had basal polytomies and satisfied the relative size condition. The resulting two-introduction likelihood was  $P((\tau_P,\tau_P) | I_2) = 2.5\%$. 

I then calculated a Bayes factor using Equation \ref{eq:bayes_factor_detailed}, the newly calculated likelihoods $P(\tau_{1C} | I_1)$, $P(\tau_{2C} | I_1)$ and $P((\tau_P,\tau_P) | I_2)$, the authors' posterior probabilities, and the compatibility equations \ref{eq:compatibility_pp}, \ref{eq:compatibility_1c}, and \ref{eq:compatibility_2c}. The results are compared to the authors' reported results in Table \ref{tab:assume}. The two-introduction likelihood is reduced by an order of magnitude. The Bayes factor is reduced to below 0.5, reversing direction against the report's conclusions. 

\begin{table}[htbp]
    \centering
    \begin{tabular}{lcc}
    \toprule
    & \textbf{Reported} & \textbf{Re-analysis with relative size condition} \\
    \midrule
    Failure rate & 77.8\% & 78.4\% \\
    \(P(\tau_{1C} | I_1)\) & 3.1\% & 3.1\% \\
    \(P(\tau_{2C} | I_1)\) & 0.0\% & 0.1\% \\
    \(P((\tau_P,\tau_P)|I_2)\) & 22.6\% & 2.5\% \\
    BF & 4.3 & 0.5 \\
    \bottomrule
    \end{tabular}\\[1ex]
    \centering
    \caption{Comparison of results as reported and from re-analysis balancing the relative size condition.}
    \label{tab:assume}
\end{table}

\section{Exploring the effect of the separation and root shape conditions}

For two introductions, the likelihood of satisfying the  evolutionary-separation and root-shape conditions depends on the upstream diversity in the reservoir from which the two lineages are introduced. 

The authors did not attempt to model the upstream diversity. However, the authors do not propose any selection mechanisms in the reservoir or the introduction process. Absent explicit selection, the default assumption is neutrality, i.e. that each pair of lineages evolves from an MRCA, randomly accruing mutations over the time ($t$) in a Poisson process. For the purposes of this analysis, I assume the rate of the Poisson process is the same as that used by the authors in humans. Thus, I assume a strict molecular clock across humans and the reservoir.

\begin{figure}[htbp]
    \centering
    \includegraphics[width=1\linewidth]{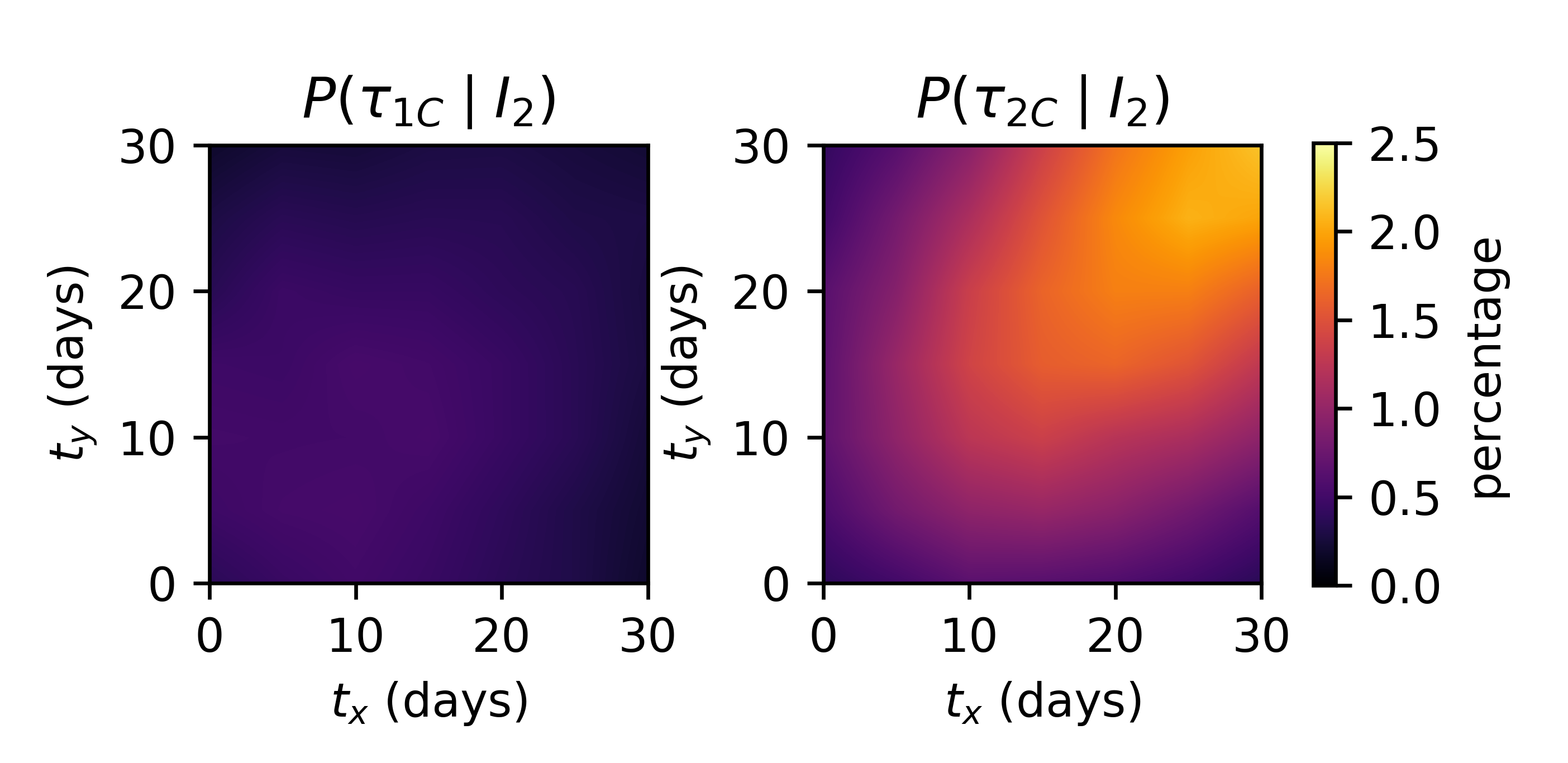}
                \caption{Two-introduction likelihoods for introduction timings $(t_x, t_y)$}
    \label{fig:two_intro}
\end{figure}

In order to explore the two-introduction likelihoods under these assumptions, I drew from the 110,000 successful epidemics, with replacement, to obtain 110,000 pairs of successful epidemics. I combined each pair for a range of times $(t_x, t_y)$ between the MRCA and the respective introductions, specifically all pairs of times $(t_x, t_y) \in \{0,5,10,15,20,25,30\}^2$ days, to obtain 110,000 arrays of 49 combined epidemics. Then, following the authors' protocols, I simulated sequencing, coalescence and evolution to generate a virus tree from each combined epidemic, and counted the proportion with topologies $\tau_{1C}$ and $\tau_{2C}$, where the two clades arose from the separate introductions. The resulting two-introduction likelihoods are shown in Figure \ref{fig:two_intro}.

Notably, Figure \ref{fig:two_intro} shows how the conditions combine to place conflicting demands on $P(\tau_{1C} | I_2)$ and $P(\tau_{2C}| I_2)$. The relative size condition rewards simultaneous introductions, i.e. $t_x = t_y$. The evolutionary separation condition rewards a large total time between the MRCA and introductions, i.e. $(t_x + t_y) \to \infty$. In combination, they reward increasing timings along the diagonal $t_x = t_y$. This coincidentally reduces the frequency of clades being ancestral and therefore increases the proportion of topologies deemed $\tau_{2C}$ by the root-shape condition. Accordingly, Figure \ref{fig:two_intro} shows the likelihood $P(\tau_{2C}| I_2)$ increasing along the diagonal $t_x = t_y$. However, for a topology to be deemed $\tau_{1C}$ by the root shape condition, one of the clades must be ancestral. Figure  \ref{fig:two_intro} shows that this pulls $P(\tau_{1C}| I_2)$ back to a lower region near the origin, where the evolutionary separation condition is less likely to be satisfied. This reflects a straightforward reality; to produce two clades of similar sizes, separated by two or more mutations, with one of the clades being ancestral, requires a relatively unusual accident of growth or evolution. To put it another way, to obtain a high likelihood of obtaining two clades of similar sizes, separated by two or more mutations, the reservoir must be assumed to be deep, but to obtain a high likelihood of one of the two introductions having the ancestral haplotype, the reservoir must be assumed to be shallow. Where the authors omitted the relative size and evolutionary separation conditions, and replaced the root shape condition with simple averaging, they bypassed this conflict.

\begin{figure}[htbp]
    \centering
    \includegraphics[width=0.8\linewidth]{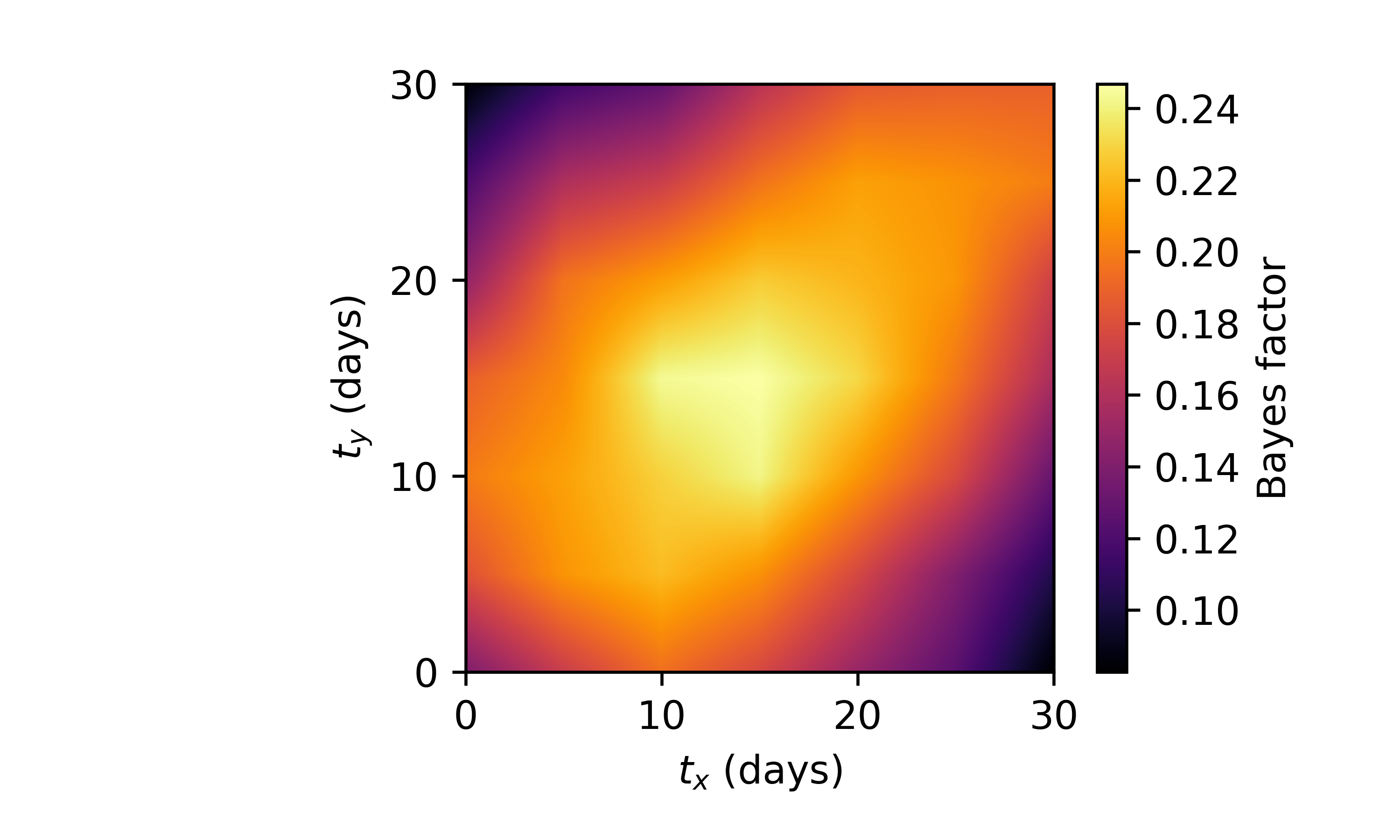}
    \caption{Bayes factors for introduction timings $(t_x,t_y)$} 
    \label{fig:bf-surfaces}
\end{figure}

I then calculated Bayes factors using Equation \ref{eq:bayes_factor_detailed}, the new likelihoods, the authors' posterior probabilities and the compatibility equations \ref{eq:compatibility_2c} and \ref{eq:compatibility_1c}. The resulting Bayes factors are shown in Figure \ref{fig:bf-surfaces}.

Notably, Figure \ref{fig:bf-surfaces} shows a maximum Bayes factor at $t_x =t_y = 15$ days, close to the timings for the maximum likelihood $P(\tau_{1C}| I_2)$ shown in Figure \ref{fig:two_intro}. This reflects the fact that the topology $\tau_{1C}$ is compatible with the more credible ancestral haplotype $S_A$, so the likelihood $P(\tau_{1C}| I_2)$ is given greater weight in the Bayes factor. The maximum Bayes factor was 0.25. The surface is also fairly flat, reflecting the noise of evolutionary processes at this scale and suggesting that a model or marginalization of $(t_x ,\, t_y)$ should produce a Bayes factor around $0.2$.

Additionally, the likelihood $P(\tau_{2C}| I_2)$ does not have a maximum inside the observed range. In order to check for a higher Bayes factor elsewhere, I calculated Bayes factors with the likelihood $P(\tau_{2C}| I_2)$ maximized at $(t_x,t_y) = (\infty,\infty)$ and the likelihood $P(\tau_{1C}| I_2)$ taken from $(t_x,t_y) = (30, 30)$, as an upper bound. The result ($BF = 0.205$) indicates that $P(\tau_{2C}| I_2)$ cannot cause a higher Bayes factor elsewhere. This reflects the lower weight the Bayes factor gives to $P(\tau_{2C}| I_2)$ for being compatible with the less credible ancestral haplotypes ${S_C,S_T}$.

Incidentally, if the two clades are not required to arise from the two introductions, the Bayes factor is trivially maximized at $(t_x,t_y) = (0,0)$, i.e. simultaneous introduction of identical viruses, so that the combined behavior is similar to a single introduction. However, this is not relevant to the hypothesis that the two clades arose from separate introductions.

The results using the strict molecular clock are compared to the authors' reported results in Table \ref{tab:results}. The one-introduction likelihoods are reproduced accurately. The two-introduction likelihoods are reduced by an order of magnitude. The maximum Bayes factor is reduced to 0.25, reversing direction to indicate at least moderate support against the report's conclusions. 

\begin{table}[htbp]
    \centering
    \begin{tabular}{lcc}
    \toprule
    & \textbf{As reported} & \textbf{Re-analysis with the strict clock} \\
    \midrule
    Failure rate & 77.8\% & 78.4\% \\
    \(P(\tau_{1C} | I_1)\) & 3.1\% & 3.1\% \\
    \(P(\tau_{2C} | I_1)\) & 0.0\% & 0.1\% \\
    \(P(\tau_{1C} | I_2)\) & 11.3\%* & 0.5\%** \\
    \(P(\tau_{2C} | I_2)\) & 11.3\%* & 1.6\%** \\
    BF & 4.3 & <0.25 \\
    \bottomrule
    \end{tabular}\\[1ex]
    \centering
    {\footnotesize
    * equivalent portion of $P((\tau_P,\,\tau_P) | I_2)$\\
    ** at $(t_x,\,t_y) = (15,\,15)$ days, for max. BF
    }
    \caption{Comparison of results as reported and from re-analysis with the strict molecular clock.}
    \label{tab:results}
\end{table}

As this model assumes neutrality and the strict molecular clock, the upstream reservoir influences evolution solely by the times $(t_x, t_y)$ between the MRCA and each introduction. The independence of each introduction also means that the relative sizes are affected solely through a difference $| t_x-t_y|$ in the introduction times. 

A biological reservoir model would induce a distribution $\pi(t_x,t_y)$ over the $(t_x,t_y)$ surface. Since the one-introduction likelihood is unchanged, the Bayes factor under any such reservoir model is

\begin{equation}
\begin{aligned}
\mathrm{BF}_{\pi}
&=
\frac{\int P(\mathbf{Y}\mid I_2,t_x,t_y)\,\pi(t_x,t_y)\,dt_x\,dt_y}{P(\mathbf{Y}\mid I_1)} \\
&=
\int \mathrm{BF}(t_x,t_y)\,\pi(t_x,t_y)\,dt_x\,dt_y
\le
\max_{t_x,t_y}\mathrm{BF}(t_x,t_y).
\end{aligned}
\label{eq:bf_upper_bound}
\end{equation}

Thus, under neutrality and the strict molecular clock, upstream reservoir size or structure cannot increase the Bayes factor above the maximum reported here. 

\section{Summary}

The authors imposed three additional conditions on the one-introduction model. Two were simple constraints. The other determined weighting.

When one of the constraints was applied to the two-introduction model, the Bayes factor was reduced from 4.3 to 0.5. 

When upstream evolution was modeled under neutrality, with a strict molecular clock comparable to that observed in humans, so that all three additional conditions could be applied to the two-introduction model, an upper bound for the Bayes factor was reduced to 0.25. 

When the conditions applied to the two-introduction model are made consistent with those that the authors applied to the one-introduction model, the Bayes factor is reduced by an order of magnitude,  and the previously reported quantitative support is reversed. The credibility of the multiple introduction hypothesis should be re-examined in light of this reversal.

\section{Methods}

\textit{Simulation of epidemics} followed the authors' protocol. For each attempted introduction, a contact network of 5 million nodes was generated with the Barabási–Albert algorithm \cite{barabasi1999emergence}, starting from 8 unconnected nodes, adding 8 edges with each new node, using preferential attachment and retrying self- and duplicate-edges. Disease progression was modeled using seven states: susceptible ($S$); exposed ($E$); pre-symptomatic and infectious ($P$); symptomatic, infectious and unascertained ($U$); symptomatic, infectious and ascertained ($A_1$); symptomatic, infectious, ascertained and proceeding to hospitalization ($A_2$); hospitalized ($H$); and recovered ($R$). The standard disease progression was 
\[
S\rightarrow E\rightarrow P \rightarrow U \rightarrow R. 
\]
However, a fraction of cases became ascertained, progressing as
\[
S\rightarrow E\rightarrow P \rightarrow A_1 \rightarrow R.
\]
A fraction of ascertained cases also became hospitalised, progressing as 
\[
S\rightarrow E\rightarrow P \rightarrow A_1 \rightarrow A_2 \rightarrow H \rightarrow R.
\]

Nodes in the infectious states $P$, $U$, $A_1$ and $A_2$ transmitted to neighbouring nodes in the susceptible $S$ state. The transmission rate for the pre-symptomatic and unascertained states $P$ and $U$ was a fraction of that for the ascertained states $A_1$ and $A_2$.  All processes were modeled as Poisson. Parameters for the epidemic simulations, and their sources, are in Table \ref{tab:simulations}.

When a susceptible node became exposed with more than one infectious neighbour, a single source was randomly selected from among the infectious neighbours, with odds weighted according to transmission rates.

Each attempted introduction was started by initializing a single randomly selected node in the exposed $E$ state, with the rest susceptible $S$, and run for 100 days. An introduction was deemed successful if, at the end of the simulation, at least 400 infections had occurred and at least one infection was not recovered. Additionally, I always deemed an introduction successful if it reached 50,000 infections (I considered the likelihood of the epidemic dying out within the 100 days after reaching 50,000 infections to be negligible). I repeatedly attempted introductions to generate epidemics for 110,000 successful introductions. 

\begin{table}[htbp]
    \centering
    \begin{tabular}{lcc}
    \toprule
   Parameter & Value & Source \\
    \midrule
    Mean number of contacts
        & 16 
        & \cite{mossong2008social} \\
    Mean duration in $E$
        & 2.9 days 
        & \cite{he2020temporal}, \cite{li2020early}\\
    Mean duration in $P$
        & 2.3 days 
        & \cite{he2020temporal},  \cite{li2020early}\\
    Mean duration in $U$
        & 2.9 days 
        & \cite{hao2020reconstruction} \\
    Mean duration in $A_1$
        & 2.9 days
        & \cite{hao2020reconstruction} \\   
    Mean duration in $A_2$
        & 8.1 days 
        & \cite{pan2020association} \\    
    Mean duration in $H$
        & 30 days
        & \cite{hao2020reconstruction} \\
    Fraction ascertained
        & 0.15
        & \cite{hao2020reconstruction}\\
    Fraction of ascertained hospitalized
        & 0.5
        & \cite{doi:10.1126/science.abp8337}\\
    Transmission rate of ascertained
        & 0.0175 /contact/day
        & \cite{doi:10.1126/science.abp8337}\\
    Ratio of transmission rate for 
        & 0.55
        & \cite{li2020substantial}\\
    unascertained over ascertained &&\\
    \bottomrule
    \end{tabular}
    \caption{Parameters for the epidemic simulations.}
    \label{tab:simulations}
\end{table}

Each successful simulated epidemic produced a transmission network, detailing the progression of the epidemic across the network, and a list of disease events, detailing the progression of the disease in each infected host.

Notably, although the simulations followed the protocol described by the authors, they differed from what was actually implemented by the authors. The authors hard coded their script to ignore commands to initialize the primary case in the exposed state ($E$) and instead initialized the primary case in the pre-symptomatic and infectious state ($P$). I assumed the protocol and command indicated the intended behaviour and therefore initialized the primary case in the exposed state ($E$).

\textit{Combining epidemics} did not follow the authors' protocol. The authors merely assumed that each introduction was independent, and calculated $P((\tau_P,\,\tau_P) | I_2)$ as $P(\tau_P | I_1)^2$. Instead, I drew from the 110,000 successful epidemics, with replacement, to obtain 110,000 pairs of successful epidemics $(E_x,\,E_y)$. For each pair of introductions, I made 49 unique combinations with timings from $(t_x, t_y) \in \{0,5,10,15,20,25,30\}^2$. 

For each combined transmission network, I added an MRCA that immediately divided into two lineages, and made the two lineages transmit to the primary cases of the epidemics $E_x$ and $E_y$ at $t_x$ and $t_y$, respectively. I also offset the timings of the transmissions in $E_x$ and $E_y$ by $t_x$ and $t_y$, respectively, and merged them into the combined transmission network. I stopped the merge once 50,000 transmissions had been included, or once one of the two simulations had reached its end time - whichever came first.

For each combined list of disease events, I offset the timings of the disease events in $E_x$ and $E_y$ by $t_x$ and $t_y$, respectively, and merged them into a combined disease events list. I only included disease events for cases included in the combined transmission network, and I stopped the merge once one of the two simulations had reached its end time.

\textit{Simulation of sequencing} followed the authors' protocol. For each ascertained case among the  first 50,000 infections of an epidemic or combined epidemic, a notional sequence was sampled at a time randomly selected from a period starting from symptom onset (i.e. $P \rightarrow A_1$) and ending at recovery (i.e. $A_1 \rightarrow R$ or $H \rightarrow R$). Sequences with sampling times preceding the first hospitalization (i.e. the first $A_2 \rightarrow H$) were discarded. If the start of the period for a hospitalized case preceded the first hospitalization, it was set to the time of the first hospitalization, in order to ensure that all hospitalized cases were sequenced. If an ascertained case had not recovered by the end of the simulation, the end of the period was set to the end of the simulation.

\textit{Simulation of coalescence} followed the authors' protocol. For each epidemic and combined epidemic, the ancestry of the simulated virus samples was traced back through the transmission network to produce a virus phylogeny. Within hosts, lineages were merged with Kingman's coalescent under a fixed effective population of one year. For events between a host's earliest transmission and infection, the exponential distribution used to sample waiting times was truncated, in order to ensure a single lineage at the time of infection. 

\textit{Simulation of evolution} followed the authors' protocol. Mutations were approximated as unique and irreversible, and simulated along the virus phylogeny as a Poisson process, at a rate of one mutation every 13.27 days. For $(t_x,t_y) = (\infty,\infty)$ the simulations started from the respective clade roots following each introduction. Note that the likelihood when $(t_x,t_y) = (\infty,\infty)$ corresponds to $P((t_p,t_p)| I_2)$.

\textit{Clade analysis} followed the authors' protocol. Lineage numbers included all leaves and mutated internal nodes descending directly from a clade root (i.e. without intermediate mutated nodes). Clade sizes counted all leaves descending directly or indirectly from clade roots.

\textit{Checking if two clades arose from two introductions} was performed by running the clade analysis for two-introduction trees with the clade roots manually set to the first internal nodes following each introduction.

\textit{Stable coalescence} was not used. The stable coalescence effectively prunes short-lived basal lineages. Therefore, it might increase the one-introduction likelihoods, but should have no effect on the two-introduction likelihoods, because the basal lineages of the two introductions always succeed, as a condition of Equation \ref{eq:bayes_factor}. However, the authors did not explain the reason for using the stable coalescence. Therefore, I concluded that using the stable coalescence could only benefit the one-introduction case and decided to omit it. The one-introduction likelihoods nonetheless almost perfectly matched those of the authors, suggesting that the effect of the stable coalescence is less than the sampling error, and that the effect of the omission was negligible.  

\section{Data and reproducibility}

Results and code are available at \href{https://github.com/nizzaneela/re-multi-introductions}{https://github.com/nizzaneela/re-multi-introductions}. 

The results include the clade analysis result for each simulated virus tree along with a hash of the corresponding tree.

The code includes a script for reproducing all simulations and a notebook for collating the results. A separate notebook provides verification of the clade analysis results and hashes from arbitrary simulations.

The scripts simulate the epidemics using a modified version of Futing Fan's implementation \cite{fang2019improving} of GEMF \cite{darabi2013generalized} \cite{SAHNEH201736}, available at \href{https://github.com/nizzaneela/GEMF}{https://github.com/nizzaneela/GEMF}. The modified version of GEMF includes Niema Moshiri's modifications to read a seed for the pseudo-random number generator from the input file.

The scripts use CoaTran \cite{Moshiri2020.11.10.377499} for coalescence and TreeSwift \cite{moshiri2020treeswift} for the clade analysis.

%Bibliography
\bibliographystyle{unsrt}  
\bibliography{references}  

\end{document}